\documentclass[12pt]{article}

\usepackage{amsfonts}
\usepackage[centertags]{amsmath}
\usepackage{amssymb}
\usepackage{cite}
\usepackage{psfrag}
\usepackage{graphicx}
\usepackage{bbm,bm}
\usepackage{epsfig, multicol}

\allowdisplaybreaks
\begin{document}

\topmargin 0pt
\oddsidemargin 0mm
\def\be{\begin{equation}}
\def\ee{\end{equation}}
\def\bea{\begin{eqnarray}}
\def\eea{\end{eqnarray}}
\def\ba{\begin{array}}
\def\ea{\end{array}}
\def\ben{\begin{enumerate}}
\def\een{\end{enumerate}}
\def\nab{\bigtriangledown}
\def\tpi{\tilde\Phi}
\def\nnu{\nonumber}
\newcommand{\eqn}[1]{(\ref{#1})}

\newcommand{\half}{{\frac{1}{2}}}
\newcommand{\vs}[1]{\vspace{#1 mm}}
\newcommand{\dsl}{\pa \kern-0.5em /} 
\def\a{\alpha}
\def\b{\beta}
\def\g{\gamma}\def\G{\Gamma}
\def\d{\delta}\def\D{\Delta}
\def\ep{\epsilon}
\def\et{\eta}
\def\z{\zeta}
\def\t{\theta}\def\T{\Theta}
\def\l{\lambda}\def\L{\Lambda}
\def\m{\mu}
\def\f{\phi}\def\F{\Phi}
\def\n{\nu}
\def\p{\psi}\def\P{\Psi}
\def\r{\rho}
\def\s{\sigma}\def\S{\Sigma}
\def\ta{\tau}
\def\x{\chi}
\def\o{\omega}\def\O{\Omega}
\def\k{\kappa}
\def\pa {\partial}
\def\ov{\over}
\def\nn{\nonumber\\}
\def\ud{\underline}
\begin{flushright}
%
\end{flushright}
\begin{center}
{\large{\bf Conformally de Sitter space from anisotropic\\ SD3-brane     
of type IIB string theory}}

\vs{10}

{Shibaji Roy\footnote{E-mail: shibaji.roy@saha.ac.in}}

\vs{4}

{\it Saha Institute of Nuclear Physics\\
1/AF Bidhannagar, Calcutta 700064, India\\}

\end{center}

\vs{15}

\begin{abstract}
We construct a four dimensional de Sitter space upto a conformal transformation
by compactifying the anisotropic SD3-brane solution of type IIB string theory on 
a six dimensional product space of the form $H_5 \times S^1$, where $H_5$ is 
a five dimensional hyperbolic space and $S^1$ is a circle. The radius of the 
hyperbolic space is chosen to be constant. The radius of the circle and the dilaton in four dimensions 
are time dependent and not constant in general. By different choices of
parameters characterizing the SD3-brane solution either the dilaton or the radius of the circle
can be made constant but not both. The form field is also non vanishing in general, but it can be 
made to vanish without affecting the solution. This construction might be useful for a better
understanding of dS/CFT correspondence as well as for cosmology.  
\end{abstract}

\newpage

Space-like branes \cite{Gutperle:2002ai, Maloney:2003ck} or S-branes, for short, are interesting time 
dependent Euclidean brane solutions
that exist in many field theories as well as string/M theory. These are soliton-like objects localized
in space-time and therefore exist only for a moment in time. In string theory, S-branes arise as the time
dependent solution of world-volume tachyon (or the rolling tachyon) of the non-BPS D-brane or 
D-brane/anti-D-brane pair \cite{Sen:1999mg, Sen:2002nu}. The original motivation for studying S-branes was to 
understand the holography
in the temporal context. In particular, as the spatial holography or AdS/CFT gives us an additional space-like
direction from a Lorentzian field theory, the temporal holography can similarly give a time-like direction
from an Euclidean field theory which is necessary for the dS/CFT correspondence \cite{Strominger:2001pn}. 
The space-time construction
of these S-branes was thus motivated to understand the proposed dS/CFT correspondence. 

Even though much work has been done on S-branes, the original motivation how to relate these objects to de Sitter 
space remained unclear. In this Letter we give a simple construction of four dimensional de Sitter space upto
a conformal transformation from certain S-brane solution of type IIB string theory. This not only clarifies the
relation between S-branes and de Sitter space, but also could be useful in the cosmological context. It is well-known
that our universe is going through an accelerated expansion \cite{Riess:2001gk} in the present epoch 
\cite{Lewis:2002ah, Bennett:2003bz} and S-branes \cite{Gutperle:2002ai, Chen:2002yq, Kruczenski:2002ap, Roy:2002ik, 
Bhattacharya:2003sh} are known to give rise to such
an accelerated expansion \cite{Townsend:2003fx, Ohta:2003pu, Roy:2003nd, Emparan:2003gg, Chen:2003dca} 
but the acceleration in that context is found to be transient with e-folding of the order of unity.
But if S-brane can be related conformally to de Sitter space, we can get a realistic cosmology in that particular
conformal frame. Of course, the significance of that conformal frame, which is not clear to us at present, needs to be 
understood.                

We must mention that despite the no-go theorem of Maldacena-Nunez \cite{Maldacena:2000mw}, many metastable de Sitter 
vacua have been found (circumventing 
the no-go theorem) in
the literature from Type IIB, IIA and heterotic string theory starting from the original work by Kachru-Kallosh-Linde-Trivedi 
\cite{Kachru:2003aw} by the so-called geometric/non-geometric flux compactification \cite{Dasgupta:1999ss,Giddings:2001yu}. This 
approach 
also solves another important problem of fixing 
the moduli. However, explicit classical solutions have not been reported in the literature. In this Letter we will construct one
such classical solution, which is not exactly de Sitter but conformal to it, from type IIB string theory. To be precise, we start
from the known \cite{Bai:2006vv, Lu:2007bu} non-supersymmetric configuration of intersecting charged D3-brane with 
chargeless D0-brane solution of type IIB
string theory and take a double Wick rotation to obtain the SD3-brane solution, which is anisotropic in one spatial direction.
(Note that the S-brane here is called SD-brane because it possesses the same charge as the D-brane.) This solution is characterized
by four independent parameters. We then write the solution in a convenient coordinate and take a `near horizon' limit after fixing
one of the parameters. The resultant solution can then be compactified on a five dimensional hyperbolic space \cite{Kaloper:2000jb,
Starkman:2000dy, Starkman:2001xu} of constant radius
and a circle of time dependent radius (in general) to obtain a four dimensional space, which can be shown to be conformal to a 
one-parameter family (the other two parameters can be set to some convenient values without any loss of generality) of de Sitter space. 
In this
solution the dilaton is non-constant. The free parameter of the solution can be chosen in  appropriate ways to make either
the radius of S$^1$ or the dilaton constant. But both of them can not be made constant simultaneously. The form-field also is non-zero 
in general, but it can be put to zero without affecting the solution.

The non-supersymmetric charged D$p$-brane intersecting with chargeless D0-brane solution of type II string theory is given in 
\cite{Bai:2006vv, Lu:2007bu}.
For $p=3$, it is a type IIB string theory solution and is given by,
\bea\label{d3d0}
ds^2 &=& - F^{-\half} 
\left(\frac{H}{\tilde H}\right)^{\frac{3\delta_1}{8} + \frac{3}{2}\delta_0} dt^2 + F^{-\half} 
\left(\frac{H}{\tilde H}\right)^{-\frac{5\delta_1}{8} - \half \delta_0}     
\sum_{i=1}^3 (dx^i)^2\nn 
& & \qquad\qquad\qquad\qquad\qquad + F^{\half} (H\tilde H)^{\half}\left(\frac{H}{\tilde H}\right)^{\frac
{3\delta_1}{8}}\left(dr^2 + r^2 d\Omega_5^2\right)\nn
e^{2(\phi-\phi_0)} &=& \left(\frac{H}{\tilde H}\right)^{\frac{\delta_1}{2} - 6\delta_0}, \qquad F_5 = (1 + \ast) Q {\rm Vol}
(\Omega_5)       
\eea
where the various functions appearing in the above solution are defined as,
\bea\label{functions}
H &=& 1 + \frac{\omega^4}{r^4}, \qquad \tilde H = 1 - \frac{\omega^4}{r^4}\nn
F &=& \left(\frac{H}{\tilde H}\right)^\alpha \cosh^2\theta - \left(\frac{\tilde H}{H}\right)^\beta \sinh^2\theta.
\eea
In \eqref{d3d0} $\phi$ is the dilaton field and $\phi_0$ is its asymptotic value. $F_5$ is the self-dual five form, 
where $Q$ is the charge parameter
and $\ast$ denotes the Hodge dual.
The solution is characterized by seven parameters, namely, $\a,\,\b,\,\d_0,\,\d_1,\,\omega,\,\theta,$ and $Q$. The equations of motion
give us three relations among them and they are
\bea\label{relations}
& & \a-\b = -\frac{3}{2}\d_1\nn
& & \half\d_1^2 + \half \a(\a+\frac{3}{2}\d_1) + \frac{3}{2}\d_0^2 = (1-3\d_0^2)\frac{5}{4}\nn
& & Q =  4 \omega^4 (\a+\b)\sinh2\theta.
\eea
Therefore, there are actually four independent parameters in the solution. Now we can get the anisotropic 
space-like D3 brane or SD3-brane solution
from the above solution \eqref{d3d0} by the double Wick rotation $r \to it$, $t \to - ix^4$, alongwith 
$\omega \to i\omega$, $\theta_1 \to i\theta_1$
(where $\theta_1$ is one of the angles of the sphere $\Omega_5$) and $\theta \to i\theta$. The solution 
\eqref{d3d0} then takes the form,
\bea\label{sd3}
ds^2 &=&  F^{\half} (H\tilde H)^{\half}\left(\frac{H}{\tilde H}\right)^{\frac{3\delta_1}{8}}\left(-dt^2 + t^2 dH_5^2\right)
+ F^{-\half} 
\left(\frac{H}{\tilde H}\right)^{\frac{3\delta_1}{8} + \frac{3}{2}\delta_0} (dx^4)^2\nn 
& & \qquad\qquad\qquad\qquad\qquad + F^{-\half} \left(\frac{H}{\tilde H}\right)^{-\frac{5\delta_1}{8} - \half \delta_0}     
\sum_{i=1}^3 (dx^i)^2\nn 
e^{2(\phi-\phi_0)} &=& \left(\frac{H}{\tilde H}\right)^{\frac{\delta_1}{2} - 6\delta_0}, \qquad F_5 = -(1 + \ast) Q {\rm Vol}(H_5),       
\eea
where the various functions are now defined as,
\bea\label{newfunctions}
H &=& 1 + \frac{\omega^4}{t^4}, \qquad \tilde H = 1 - \frac{\omega^4}{t^4}\nn
F &=& \left(\frac{H}{\tilde H}\right)^\alpha \cos^2\theta + \left(\frac{\tilde H}{H}\right)^\beta \sin^2\theta.
\eea  
The parameter relations \eqref{relations} remain exactly the same except the last one, which gets modified as
$Q = 4 \omega^4 (\a+\b)\sin 2\theta$. Eq.\eqref{sd3} represents the anisotropic SD3-brane solution which has
a four dimensional Euclidean world-volume with one anisotropic direction $x^4$. Also note that the solution
is well-defined as long as $t > \omega$.

We now make a coordinate transformation,
\be\label{coord}
t = \tau \left(\frac{1 + \sqrt{g(\tau)}}{2}\right)^{\half}, \qquad {\rm where}\quad g(\tau) = 1 + \frac{4\omega^4}{\tau^4} \equiv
1 + \frac{\tau_0^4}{\tau^4}.
\ee
In terms of this new coordinate the metric and the dilaton in \eqref{sd3} take the forms,
\bea\label{solninnewcoord}
ds^2 &=& G^{\half} g^{\frac{\a}{4} + \frac{1}{4} + \frac{3\d_1}{16}}\left(-\frac{d\tau^2}{g}+\tau^2 dH_5^2\right) + G^{-\half}
 g^{-\frac{\a}{4} + \frac{3\d_1}{16} + \frac{3\d_0}{4}} (dx^4)^2\nn
& &\qquad\qquad\qquad\qquad\qquad + G^{-\half}  g^{-\frac{\a}{4} - \frac{5\d_1}{16} - \frac{\d_0}{4}} \sum_{i=1}^3 (dx^i)^2\nn
e^{2(\phi - \phi_0)} &=& g^{\frac{\d_1}{4} - 3\d_0}, \qquad {\rm where,}\quad G\,\,=\,\, \cos^2\theta + g^{-1}\sin^2\theta.
\eea 
In writing the function $G$, we have put $\a+\b=2$ for simplicity as we will see. We also have from \eqref{relations},
$\a-\b=-(3/2)\d_1$ and so, combining these two we get, $\a = 1-(3/4)\d_1$. Using this in the second parameter relation,
given in \eqref{relations}, we get
\be\label{parameterrelations}
\frac{7}{8}\d_1^2 + 21 \d_0^2 = 3.
\ee
Now we take the `near horizon' limit $\tau \ll \tau_0$ and in that case $g(\tau) = 1 + \tau_0^4/\tau^4 \approx \tau_0^4/\tau^4$
and $G = \cos^2\theta + g^{-1} \sin^2\theta \approx \cos^2\theta$. With these the metric and the dilaton reduce to,
\bea\label{solninnewcoord1}
& & ds^2 = -\frac{\cos\theta \tau^2}{\tau_0^2} d\tau^2 + \frac{\tau_0^{3\g - 1}}{\cos\theta \tau^{3\g- 1}} (dx^4)^2
+  \frac{\tau^{\g + 1}}{\cos\theta \tau_0^{\g + 1}}\sum_{i=1}^3 (dx^i)^2 + \cos\theta \tau_0^2 dH_5^2\nn
& & e^{2(\phi - \phi_0)} =  \frac{\tau_0^{\d_1 -12\d_0}}{\tau^{\d_1 - 12\d_0}}, \qquad {\rm where}\quad \g = \half \d_1 + \d_0.
\eea  
Note that here we have obtained the radius of the five dimensional hyperbolic space to be constant given by $\tau_0
\sqrt{\cos\theta}$. This happened because we had chosen the condition $\a+\b=2$. If we did not use this condition
the radius would have been time dependent. This simplifies the compactification on hyperbolic space. Further, for
simplification let us put $\tau_0=1$ and $\cos\theta=1$. Note that $\cos\theta=1$ implies $Q=0$ and therefore
the form-field is absent. Again this is not necessary, but we use this just for simplification. Now with these
simplifications we can compactify the metric successively on $H_5$ and $x^4$ (notice that $x^4$ has a time dependent radius
in general when $3\gamma -1 \neq 0$), 
i.e., on a product space of the type
$H_5 \times S^1$ and the resultant four dimensional metric in the Einstein frame as well as the dilaton can be written as,
\bea\label{dsmetric}
& & ds_4^2 =  \eta^3\left[-\frac{4}{(\gamma-3)^2} \frac{d\eta^2}{\eta^2} + \frac{\sum_{i=1}^3 (dx^i)^2}{\eta^2}\right]\nn
& & e^{2(\phi-\phi_0)} = \eta^{\frac{2(\d_1-12\d_0)}{\gamma-3}},
\eea
where we have defined a new time coordinate by the relation $\tau^{\gamma-3} = \eta^{-2}$. We recognize the metric in \eqref{dsmetric}
to be a one-parameter family of four dimensional de Sitter metric upto a conformal factor $\eta^3$. One may think that since
we can freely choose the parameter ($\d_1$ or $\d_0$ related by the relation \eqref{parameterrelations}), we can use the freedom
to make the metric purely de Sitter, but that is not possible. The reason is, since $\d_1$ and $\d_0$ have to be real, they can not
take arbitrary values. In fact, we see from \eqref{parameterrelations} that we must have $0\leq |\d_1| \leq 2\sqrt{6}/\sqrt{7}$, 
such that both $\d_0$ and $\d_1$ remain real. Similarly, we should have $0\leq 
|\d_0| \leq 1/\sqrt{7}$, such that the parameters remain real. Because of this restriction, we can not have a de Sitter metric from 
SD3-brane. Note that the constant factor in front of $-d\eta^2/\eta^2$ term can be scaled away by scaling $\eta$ and $x^{1,2,3}$ 
appropriately. Also, by using the parameter relation \eqref{parameterrelations} we can easily check that the maximum value of
$\gamma=(1/2)\d_1 + \d_0$ is $+1$ and the minimum value is $-1$. The maximum of $\gamma$ occurs when $\d_1 = 12/7$ and $\d_0=1/7$
and the minimum occurs when $\d_1= -12/7$ and $\d_0 = -1/7$. At these two extremal values the exponent of $\eta$ in the dilaton 
expression given in \eqref{dsmetric} vanishes and the dilaton becomes constant. The metric also simplifies at these two extremal
values of $\gamma$. Thus, it is clear that the de Sitter metric upto a conformal factor can actually be obtained from pure gravity
in ten dimensions without the dilaton and the form-field. Also note that the radius of $x^4$ can be made constant by putting
$\gamma = (1/2)\d_1+\d_0 = 1/3$. The value of $\d_0$ and $\d_1$ in that case can be determined from \eqref{parameterrelations}
to be, $\d_0 \simeq 0.38\,\, {\rm or,}\,\, -0.28$ and $\d_1 \simeq -0.09 \,\, {\rm or,} \,\, 1.23$. But,
with these values of the parameters the dilaton would be time dependent. 
We would like to mention that we could have relaxed the condition $\a+\b=2$ and in that case radius of the hyperbolic space would
also become time dependent as we mentioned before and one might think that this would give more freedom to choose the parameters
and the compactification might lead to purely de Sitter space in four dimensions. But, again we have checked that this is not
possible. The parameters are restricted in such a way that we can obtain de Sitter only upto a conformal factor. Finally, we would like to
point out that in order for the supergravity description to remain valid in various regimes of time, the radius of the circle $S^1$ must 
be large compared to the
string scale, otherwise, we have to go to the T-dual frame (or type IIA theory) and the effective string coupling 
$e^{2\phi}$ must remain small 
compared to $e^{2\phi_0} = g_s^2$ (where $g_s$ is the string coupling), otherwise, we have to go to the S-dual 
frame (of type IIB theory).  

To understand the significance of de Sitter space upto a conformal transformation in string theory SD3-branes, we point out
that this is analogous to the near horizon limit of D$p$-branes of type II string theories for $p \neq 3,5$, where 
AdS spaces show up upto a conformal transformation. The implication of the appearance of conformally AdS spaces in the
latter case is that, there exists a more general correspondence, than AdS/CFT, which has been named as domain wall/QFT
correspondence \cite{Boonstra:1998mp}. In fact, it has been shown in \cite{Boonstra:1998mp}, that there exists a special 
conformal frame (or dual frame) in
which the 10-dimensional supergravity action leads to a $p+2$-dimensional gauged supergravity action upon compactification
on S$^{8-p}$. The corresponding Einstein frame action then admits an AdS$_{p+2}$ `linear dilaton' vacuum which is equivalent
to the domain wall solution in this conformal frame. One is thus led to the conjecture that the QFT describing the
internal dynamics of large $N$ number of coincident D$p$-branes is equivalent to string theory on domain wall background
\cite{Boonstra:1998mp}.
For the case of conformally de Sitter space in SD3-brane solution we expect an Euclideanized version of domain wall/QFT-like
correspondence to hold. It would be interesting to substantiate this claim with more supporting evidence, for example, comparing the
symmetries and also to see whether similar correspondences hold for other SD-branes of type II string theories.

Thus, in this Letter, we have given a simple construction of a four dimensional de Sitter space upto a conformal factor by compactifying 
the anisotropic SD3-brane solution of type IIB string theory on a six dimensional internal space of the form $H_5 \times S^1$. The 
radius of $H_5$ is chosen to be constant. The radius of $S^1$ and the dilaton are in general time dependent. However, they can be made
constant separately by appropriate choices of parameters. This brings out the connection between space-like branes and the de Sitter
space which might be useful to understand dS/CFT correspondence better in the same spirit as AdS/CFT. 
Our construction may also be useful in the
context of cosmology, as looking from a particular conformal frame, we obtain a four dimensional universe with a 
positive cosmological constant, consistent with observation.

\vspace{.5cm}

\noindent{\it Acknowledgements:} I am grateful to J. X. Lu for very useful comments and discussions.  I would also like to thank Amit Ghosh 
and Harvendra Singh for useful discussions.

\end{document}